\documentstyle[preprint,aps,epsf,tighten]{revtex}
\begin{document}
%\draft

\preprint{Fermilab-Pub-96/430-E}

\title{Search for a Fourth Generation Charge $-1/3$ Quark via
Flavor Changing Neutral Current Decay}

%\input{list_of_authors}
% LIST_OF_AUTHORS.TEX                 10/31/96           
%
\author{                                                                        
%% names begin here                                                             
S.~Abachi,$^{14}$                                                               
B.~Abbott,$^{28}$                                                               
M.~Abolins,$^{25}$                                                              
B.S.~Acharya,$^{43}$                                                            
I.~Adam,$^{12}$                                                                 
D.L.~Adams,$^{37}$                                                              
M.~Adams,$^{17}$                                                                
S.~Ahn,$^{14}$                                                                  
H.~Aihara,$^{22}$                                                               
G.~\'{A}lvarez,$^{18}$
% ...
\\                                                                              
\vskip 0.50cm                                                                   
\centerline{(D\O\ Collaboration)}                                               
\vskip 0.50cm                                                                   
}                                                                               
\address{                                                                       
\centerline{$^{1}$Universidad de los Andes, Bogot\'{a}, Colombia}               
\centerline{$^{2}$University of Arizona, Tucson, Arizona 85721}                 
\centerline{$^{3}$Boston University, Boston, Massachusetts 02215}               
\centerline{$^{4}$Brookhaven National Laboratory, Upton, New York 11973}        
\centerline{$^{5}$Brown University, Providence, Rhode Island 02912}             
\centerline{$^{6}$Universidad de Buenos Aires, Buenos Aires, Argentina}         
\centerline{$^{7}$University of California, Davis, California 95616}            
\centerline{$^{8}$University of California, Irvine, California 92717}           
\centerline{$^{9}$University of California, Riverside, California 92521}        
\centerline{$^{10}$LAFEX, Centro Brasileiro de Pesquisas F{\'\i}sicas,          
                  Rio de Janeiro, Brazil}                                       
\centerline{$^{11}$CINVESTAV, Mexico City, Mexico}                              
\centerline{$^{12}$Columbia University, New York, New York 10027}               
\centerline{$^{13}$Delhi University, Delhi, India 110007}                       
\centerline{$^{14}$Fermi National Accelerator Laboratory, Batavia,              
                   Illinois 60510}                                              
\centerline{$^{15}$Florida State University, Tallahassee, Florida 32306}        
\centerline{$^{16}$University of Hawaii, Honolulu, Hawaii 96822}                
\centerline{$^{17}$University of Illinois at Chicago, Chicago, Illinois 60607}  
\centerline{$^{18}$Indiana University, Bloomington, Indiana 47405}              
\centerline{$^{19}$Iowa State University, Ames, Iowa 50011}                     
\centerline{$^{20}$Korea University, Seoul, Korea}                              
\centerline{$^{21}$Kyungsung University, Pusan, Korea}                          
\centerline{$^{22}$Lawrence Berkeley National Laboratory and University of      
                   California, Berkeley, California 94720}                      
\centerline{$^{23}$University of Maryland, College Park, Maryland 20742}        
\centerline{$^{24}$University of Michigan, Ann Arbor, Michigan 48109}           
\centerline{$^{25}$Michigan State University, East Lansing, Michigan 48824}     
\centerline{$^{26}$Moscow State University, Moscow, Russia}                     
\centerline{$^{27}$University of Nebraska, Lincoln, Nebraska 68588}             
\centerline{$^{28}$New York University, New York, New York 10003}               
\centerline{$^{29}$Northeastern University, Boston, Massachusetts 02115}        
\centerline{$^{30}$Northern Illinois University, DeKalb, Illinois 60115}        
\centerline{$^{31}$Northwestern University, Evanston, Illinois 60208}           
\centerline{$^{32}$University of Notre Dame, Notre Dame, Indiana 46556}         
\centerline{$^{33}$University of Oklahoma, Norman, Oklahoma 73019}              
\centerline{$^{34}$University of Panjab, Chandigarh 16-00-14, India}            
\centerline{$^{35}$Institute for High Energy Physics, 142-284 Protvino, Russia} 
\centerline{$^{36}$Purdue University, West Lafayette, Indiana 47907}            
\centerline{$^{37}$Rice University, Houston, Texas 77005}                       
\centerline{$^{38}$Universidade Estadual do Rio de Janeiro, Brazil}             
\centerline{$^{39}$University of Rochester, Rochester, New York 14627}          
\centerline{$^{40}$CEA, DAPNIA/Service de Physique des Particules, CE-SACLAY,   
                   France}                                                      
\centerline{$^{41}$Seoul National University, Seoul, Korea}                     
\centerline{$^{42}$State University of New York, Stony Brook, New York 11794}   
\centerline{$^{43}$Tata Institute of Fundamental Research,                      
                   Colaba, Bombay 400005, India}                                
\centerline{$^{44}$University of Texas, Arlington, Texas 76019}                 
\centerline{$^{45}$Texas A\&M University, College Station, Texas 77843}         
}                                                                               
%end                                                                            

\date{\today}

\maketitle

\begin{abstract}
We report on a search for pair production of a fourth generation 
charge $-1/3$ quark ($b^\prime$) in $p\bar p$ collisions 
at $\sqrt{s} = 1.8$ TeV at the Fermilab 
Tevatron using an integrated luminosity of 93 pb$^{-1}$.  Both quarks 
are assumed to decay via flavor changing neutral currents (FCNC).  The 
search uses the signatures $\gamma + 3$ jets $+ \mu$-tag 
and $2\gamma$ + 2 jets.  We see no significant excess of events over the 
expected background.  We place an upper limit on the production cross section
times branching fraction that is well below theoretical expectations for a 
$b^\prime$ decaying exclusively via FCNC for $b^\prime$ masses up to 
$m_Z + m_b$.
\end{abstract}

\pacs{}

%%%%%%%%%%%%%%%%%%%%%%%%%%%%%%%%%%%%%%%%%%%%%%%%%%%%%%%%%%%%%%%%%%%%%%%%%%%%%%%
%                           Section: Introduction                             %
%%%%%%%%%%%%%%%%%%%%%%%%%%%%%%%%%%%%%%%%%%%%%%%%%%%%%%%%%%%%%%%%%%%%%%%%%%%%%%%

%\section{Introduction}

The existence of three generations of quarks and leptons
%including
%the recently discovered top quark~\cite{top}, 
is well established in the 
Standard Model.  There is no strong expectation of additional quark and lepton 
generations in an extended Standard Model, nor are additional generations
ruled out.  Several models with new generations or arguments favoring new
generations have been presented~\cite{four}.
In this paper, we report on a search for pair production of a fourth generation 
charge $-1/3$ quark ($b^\prime$) that decays via flavor changing neutral
currents (FCNC) in $p\bar p$ collisions 
at $\sqrt{s} = 1.8$ TeV at the Fermilab 
Tevatron.
While most Standard Model FCNC processes are highly suppressed, it is
quite plausible that a light $b^\prime$ quark 
(i.e.\ $m_{b^\prime} < m_t$ and $m_{b^\prime} < m_{t^\prime}$) could decay
predominantly via FCNC if, as expected, the
charged current decay of a light $b^\prime$ quark to a  
charm quark is highly suppressed by a four-generation extension of the
CKM matrix~\cite{fcnc}.  The condition for FCNC dominance is roughly
$\left|V_{cb^\prime}/V_{t^\prime b}\right| < 10^{-2}$ to $10^{-3}$ depending
on the mass of the $b^\prime$ and $t^\prime$ quarks.
Several $e^+e^-$
collider experiments have explicitly searched for $b^\prime$ quarks decaying
via FCNC~\cite{ee}, but
until now there have been no searches for 
$b^\prime$ quarks that decay via FCNC at hadron colliders~\cite{hadron}.
The current mass limit on a $b^\prime$ quark that decays via FCNC
is the
LEP I limit of half the $Z$ boson mass~\cite{lep}.  

The data used in this search were collected with the D\O\ detector 
during the 1992--1995 Tevatron collider run and represent an
integrated luminosity of 93 pb$^{-1}$.

We assume that $b^\prime$ quarks are pair produced with the same cross section,
for a given mass, as the top quark~\cite{top_cs}.  We 
consider the signatures $b^\prime\bar{b^\prime} \to \gamma gb\bar b$
and $b^\prime\bar{b^\prime} \to \gamma \gamma b\bar b$, in which 
the photons are observed directly and $b$ quarks and gluons are observed as
hadronic jets.  In the case of 
the single photon
signature, we require that one of the $b$ quark jets have a soft muon 
tag.  We assume that
$b^\prime$ quarks decay 100\% of the time via FCNC with the relative FCNC
branching fractions determined by the Standard Model~\cite{fcnc_br}, which
are 13\% for the single photon signature and 1.6\% for the diphoton 
signature for a $b^\prime$ quark of mass 80 GeV/c$^2$.  
We have not included the three-body hadronic FCNC decay modes, such as
$b^\prime \to b q\bar q$, in the single photon 
acceptance calculation or in any quoted 
theoretical branching fractions.  
The acceptance for such modes is only slightly
lower than for the two-body decay $b^\prime \to bg$.  If three-body hadronic
decay modes were included in the acceptance calculation for the single
photon 
signature, the acceptance times branching fraction might increase by 30--50\%,
but with considerable theoretical uncertainty.
For $b^\prime$ masses above
$m_Z + m_b$, the decay channel $b^\prime \to Z + b$ is expected to dominate
other FCNC decay processes.  Thus, the sensitivity of the photon decay 
channels is
limited to $b^\prime$ masses where the $Z$ boson decay channel is not open.

%%%%%%%%%%%%%%%%%%%%%%%%%%%%%%%%%%%%%%%%%%%%%%%%%%%%%%%%%%%%%%%%%%%%%%%%%%%%%%%
%                           Section: Event Selection                          %
%%%%%%%%%%%%%%%%%%%%%%%%%%%%%%%%%%%%%%%%%%%%%%%%%%%%%%%%%%%%%%%%%%%%%%%%%%%%%%%

The D\O\ detector is described in detail in Ref.~\cite{nim}.  The detector
consists of an iron toroid muon spectrometer,
a uranium-liquid argon calorimeter, and a non-magnetic central tracking 
volume containing drift chambers, a vertex chamber, and a transition 
radiation detector.
Jets are reconstructed using a cone algorithm with radius 
${\cal R} = 0.5$ in $\eta$-$\phi$ space, where $\eta$ is 
pseudorapidity and $\phi$
is the azimuthal angle.
Muons are identified by reconstructed tracks in the muon
spectrometer.  Muons used for $b$-tagging are required to
have $p_T > 4$ GeV/c, $|\eta| < 1.1$, and to be within 
$\Delta {\cal R} < 0.5$ of
a jet axis in $\eta$-$\phi$ space.  Photon candidates are identified by the 
longitudinal and transverse shower shape of isolated 
calorimeter energy clusters, 
and by the absence of tracking chamber hits in the central tracking 
volume between the calorimeter cluster and the event vertex~\cite{photon}.
The photon isolation requirement is that the energy in an annular isolation
cone from radius 0.2 to 0.4 in $\eta$-$\phi$ space be less than 10\% of 
the photon energy.

In addition to requiring the requisite number of photons, jets, and $b$-tagging
muons, both analyses place a cut on the quantity $H_T$, which is defined
as the scalar sum of the transverse energies ($E_T$'s) of the photons, 
jets, and any 
$b$-tagging muons
in the event.  Both analyses require 
$H_T \ge 1.6\ m_{b^\prime}$.  Note that the $H_T$ cut depends explicitly on
the $b^\prime$ mass hypothesis.  The value of the $H_T$ cut is set to
maximize expected 
significance, defined as acceptance divided by the square root
of the expected background.
The cuts used by the two analyses are summarized in Table~\ref{cuts}.

%%%%%%%%%%%%%%%%%%%%%%%%%%%%%%%%%%%%%%%%%%%%%%%%%%%%%%%%%%%%%%%%%%%%%%%%%%%%%%%
%                           Section: Acceptance and Background                %
%%%%%%%%%%%%%%%%%%%%%%%%%%%%%%%%%%%%%%%%%%%%%%%%%%%%%%%%%%%%%%%%%%%%%%%%%%%%%%%

The acceptance is calculated using the {\sc herwig} event
generator~\cite{herwig} with a detector simulation based on the {\sc geant}
program~\cite{geant}.  The calculated acceptance for the two channels
is listed in Tables~\ref{sum1} and \ref{sum2}, respectively.

\begin{figure}
\epsfysize=7.5in
\centerline{\epsfbox{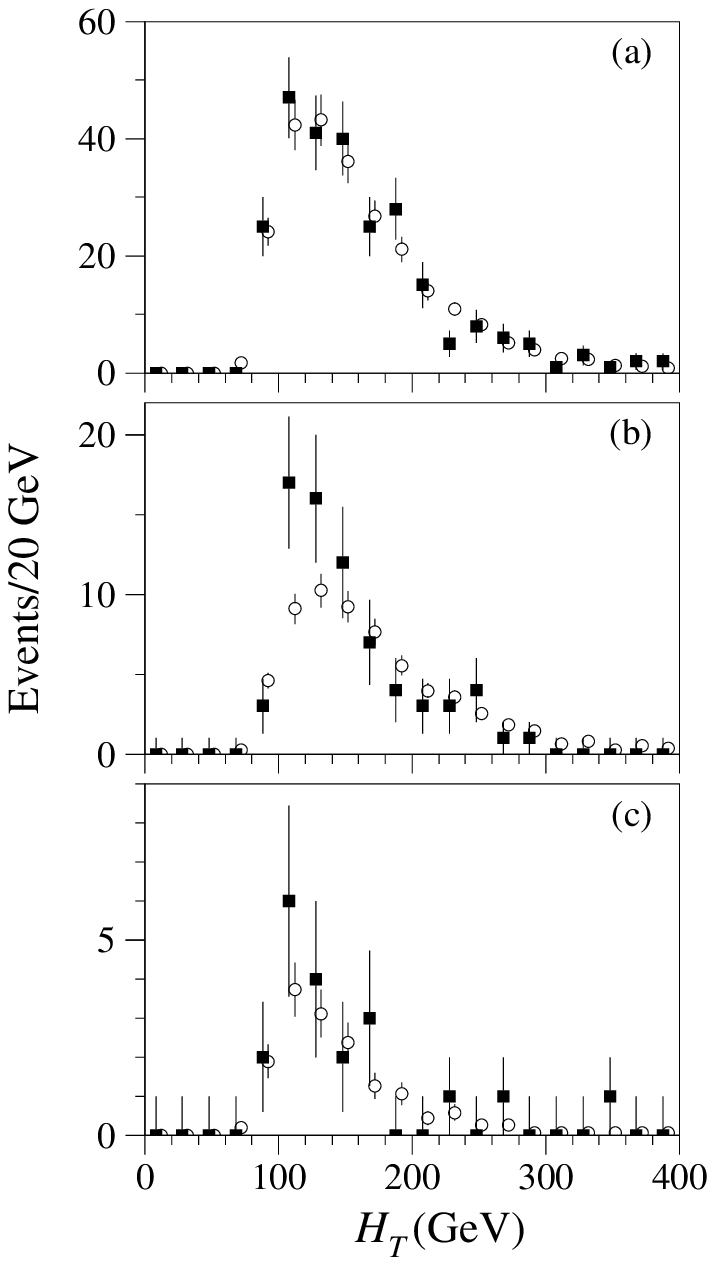}\hspace{1.4in}}
\vspace{0.6in}
\caption{\label{ht}$H_T$ distributions of data (filled squares) and 
expected background (open circles)
in the channels (a) ``bad $\gamma$'' + 3 jets, (b) $\gamma + 3$ jets, and 
(c) $2\gamma + 2$ jets.}
\end{figure}

We find 71 events before the $H_T$ cut in the single photon analysis.
The primary backgrounds to the single photon channel are QCD direct photon plus
multijet production and QCD multijet production with one jet misidentified
as a photon.  Other backgrounds that are considered are $W\gamma$ and
$Z\gamma$ production and $W$ and $Z$ bosons decaying to electron(s) with 
one electron
misidentified as a photon.  
The sum of the direct photon and multijet 
backgrounds is calculated using the tag rate method, in which untagged 
$\gamma$ + 3 jet events are weighted by a per jet 
$b$-tagging 
probability measured in multijet ($\ge4$ jet) data.  Figure~\ref{ht}(a)
shows a test of the $b$-tag rate in ``bad $\gamma$'' + 3 jets events, in which
the photon has failed one of the photon identification cuts.  There
is good agreement between data and background.
The estimated background obtained using 
the tag rate method is $62.8\pm6.3$ events before the $H_T$ cut.
The diboson background, which is expected to generate $b$-tags in excess of
the tag rate, is estimated by a Monte Carlo calculation to be 
$0.7\pm0.4$ events.  The background from $W$ and $Z$ bosons
decaying to electrons that are misidentified as photons 
is estimated to be $0.1\pm0.1$ events and is not included in the subtracted
background.  
The total expected background before the $H_T$ cut is $63.4\pm6.3$ events.
The $H_T$ distributions of data and expected background are shown in
Fig.~\ref{ht}(b).  There is a slight, but not statistically significant, 
excess of data over background.

We find 20 events before the $H_T$ cut in the diphoton channel.
The primary backgrounds to the diphoton channel are QCD multijet 
production with two jets misidentified as photons
and single direct photon plus jets production
with one jet misidentified as a photon.  Other less important
backgrounds are double direct photon + jets production and $Z\to ee$ events
where both electrons are misidentified as photons.  
The sum of the two fake
photon backgrounds is estimated from the measured probability for a jet
to be misidentified as the second photon in single photon candidate 
plus three jet 
events, corrected for the fraction of photon candidates that are actually
photons (the photon purity) of the first
photon.  The purity of the first photon in the $2\gamma + 2$ jet
sample is estimated to be $38\pm16$\%~\cite{photon}.

%Actually, the two fake backgrounds can
%be estimated separately using the photon purity of photons in 
%$\gamma + 3$ jet events, but the errors of the two backgrounds are negatively
%correlated so that the error of the sum is less than the error of either
%fake background separately.  
The sum of the two fake backgrounds is estimated
to be $14.5\pm2.2$ events before the $H_T$ cut.  
The double direct photon background is estimated
by Monte Carlo calculation to be $1.2\pm0.6$ events.  The $Z\to ee$ background
is estimated to be $0.1\pm0.1$ events and is not included in the subtracted
background.  The total expected background before the $H_T$ cut is $15.7\pm2.3$
events.  There is again a slight, but not statistically significant, excess
of data over background.  The $H_T$ distributions of data and expected
background are shown in Fig.~\ref{ht}(c).

\begin{figure}
\epsfysize=7.5in
\centerline{\epsfbox{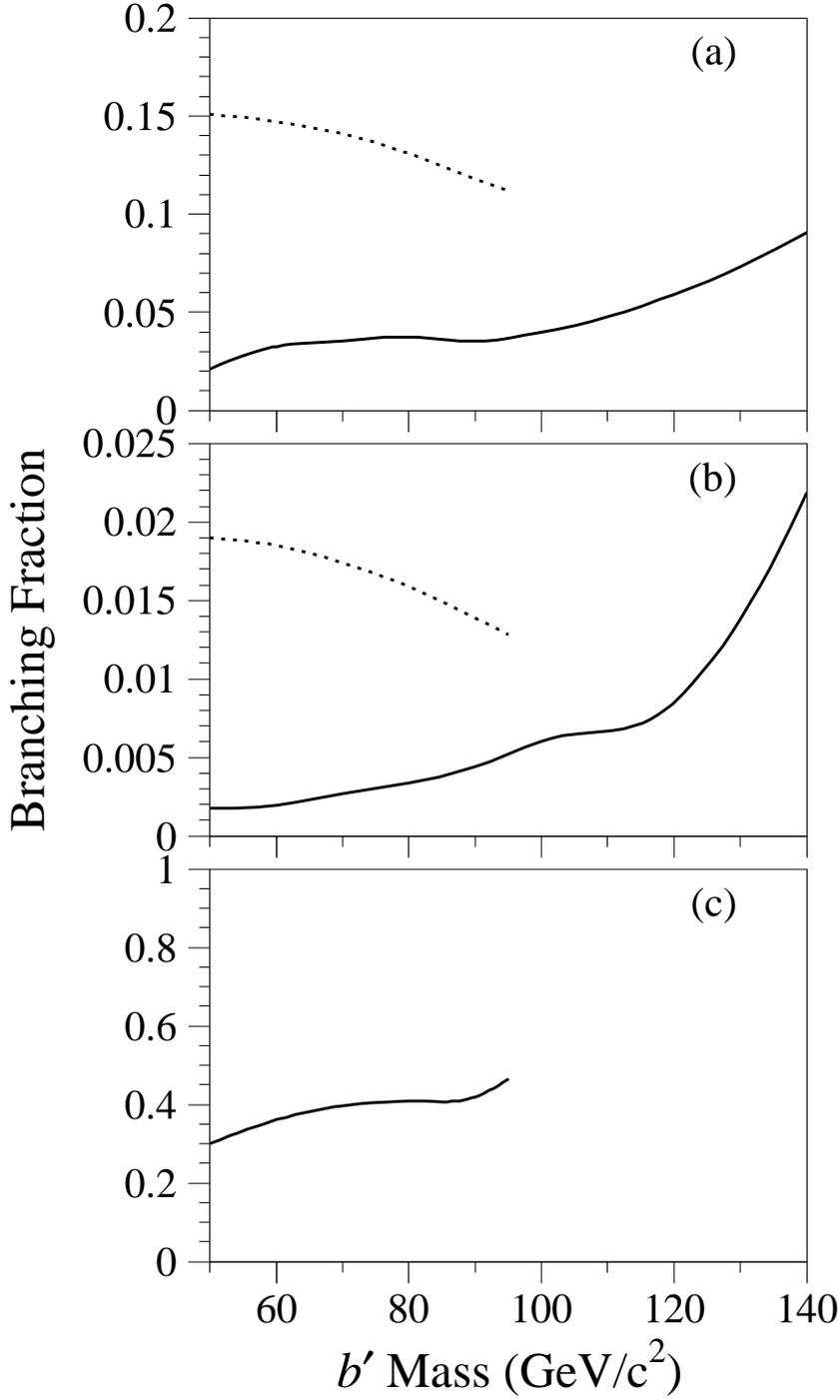}\hspace{1.4in}}
\vspace{0.6in}
\caption{\label{br}
Measured 95\% confidence level upper limit on the branching fraction 
(solid line)
and theoretical branching fraction (dotted line)
for
(a) $b^\prime\bar{b^\prime}\to\gamma + 3$ jets, 
(b) $b^\prime\bar{b^\prime}\to 2\gamma + 2$ jets, and 
(c) the total FCNC branching fraction of $b^\prime$.
The theoretical and FCNC branching fraction curves end
at $m_{b^\prime} = m_Z + m_b$ due to the opening of the $Z$ boson FCNC
decay channel.}
\end{figure}

The acceptance, the number of data events, the expected signal and 
background, including the effect of the variable $H_T$ cut, and the
calculated cross section times branching fraction are shown
in Tables~\ref{sum1} and~\ref{sum2}.  The 95\% 
confidence level (CL) upper limit on the cross section times branching fraction
is calculated using Gaussian errors excluding the unphysical
negative cross section region.
Using the theoretical production cross section
of Laenen {\it et al.}~\cite{top_cs}, including the quoted theoretical
uncertainty, we derive an upper limit on the
branching fraction for each of the two channels.  The 95\% confidence level 
upper limit for both channels is shown in Figs.~\ref{br}(a) and (b).
Using the theoretical relative FCNC branching fractions of Ref.~\cite{fcnc_br},
we derive an upper limit on the total FCNC branching fraction of the $b^\prime$
quark for both channels individually and combined.  The combined upper
limit on the FCNC branching fraction of the $b^\prime$ quark is shown in 
Fig.~\ref{br}(c).
For both channels the upper limit on the branching fraction is well
below the theoretical branching fraction for a $b^\prime$ quark that decays
100\% of the time via FCNC.  The upper limit on the total FCNC branching
fraction of the $b^\prime$ quark is 
less than 50\%, for all masses up to $m_Z + m_b$, at which point the
$Z$ boson decay channel opens up.

%\section{Acknowledgements}

%\input{acknowledgement_paragraph}
% Acknowledgement_paragraph.tex                             6/7/96
%
We thank the staffs at Fermilab and the collaborating institutions for their
contributions to the success of this work, and acknowledge support from the 
Department of Energy and National Science Foundation (U.S.A.),  
Commissariat  \` a L'Energie Atomique (France), 
Ministries for Atomic Energy and Science and Technology Policy (Russia),
CNPq (Brazil),
Departments of Atomic Energy and Science and Education (India),
Colciencias (Colombia),
CONACyT (Mexico),
Ministry of Education and KOSEF (Korea),
CONICET and UBACyT (Argentina),
and the A.P. Sloan Foundation.
%

%%%%%%%%%%%%%%%%%%%%%%%%%%%%%%%%%%%%%%%%%%%%%%%%%%%%%%%%%%%%%%%%%%%%%%%%%%%%%%%%
%                                 References                                   %
%%%%%%%%%%%%%%%%%%%%%%%%%%%%%%%%%%%%%%%%%%%%%%%%%%%%%%%%%%%%%%%%%%%%%%%%%%%%%%%%

\begin{table}
\caption{\label{cuts}Kinematic cuts used in the $\gamma$ + 3 jets and 
$2\gamma$ + 2 jets analyses (energies in GeV).}
\begin{tabular}{l|ccc|ccc|c|c}
& \multicolumn{3}{c|}{Photons} & \multicolumn{3}{c|}{Jets} & & \\
\cline{2-7}
Channel & $N$ & $E_{T\rm min}$ & $|\eta|_{\rm max}$ & 
$N_{\rm min}$ & $E_{T\rm min}$ & 
$|\eta|_{\rm max}$ & $b$-tag & $H_{T\rm min}$ \\
\hline
\hline
$\gamma$ + 3 j & 1 & $20$ & $1$ & $3$ & 15 & 2 & yes & 
$1.6\ m_{b^\prime}$ \\
$2\gamma$ + 2 j & 2 & $20$ & $2$ & $2$ & 15 & 2.5 & no & 
$1.6\ m_{b^\prime}$ \\
\end{tabular}
\end{table}

\begin{table*}
\caption{\label{sum1}The acceptance, the numbers of expected and observed 
events, and the measured cross section as a function of $b^\prime$ mass
in the $\gamma + 3$ jets channel.  The acceptance includes the muon 
semileptonic branching fraction of the $b$ quark.  The integrated luminosity
is 93 pb$^{-1}$.}
\begin{tabular}{c|c|ccc|cc}
 && \multicolumn{3}{c|}{Events} & 
   \multicolumn{2}{c}{$\sigma_{b^\prime\bar{b^\prime}}\times 
  B(b^\prime\bar b^\prime \to \gamma gb\bar b)$ (pb)}\\
\cline{3-7}
$m_{b^\prime}$  & Acceptance
&& Expected & Expected && Upper limit \\
(GeV/c$^2$) & (\%) & Observed & Signal ($B = 13$\%) & Background & 
Value & (95\% CL) \\
\hline
\hline
50 & $0.38\pm0.07$ & 71 & $166\pm33$ & $63.1\pm6.3$ & $22.1\pm30.1$ & $75.3$ \\
60 & $0.64\pm0.12$ & 70 & $115\pm22$ & $60.0\pm6.0$ & $17.0\pm17.8$ & $47.8$ \\
70 & $1.10\pm0.19$ & 60 & $87\pm16$ & $53.4\pm5.3$ & $6.5\pm9.3$ & $23.1$ \\
80 & $1.45\pm0.25$ & 46 & $57\pm10$ & $45.4\pm4.6$ & $0.4\pm6.1$ & $12.2$ \\
90 & $1.68\pm0.29$ & 30 & $35\pm6$ & $37.4\pm3.8$ & $-4.8\pm4.4$ & $6.0$ \\
100 & $2.16\pm0.36$ & 23 & $26\pm5$ & $30.1\pm3.1$ & $-3.5\pm2.9$ & $3.9$ \\
120 & $2.88\pm0.46$ & 14 & $13\pm2$ & $18.7\pm1.9$ & $-1.8\pm1.6$ & $2.2$ \\
140 & $3.50\pm0.55$ &  9 & $7\pm1$ & $12.0\pm1.3$ & $-1.0\pm1.0$ & $1.5$ \\
\end{tabular}
\end{table*}

\begin{table*}
\caption{\label{sum2}The acceptance, the numbers of expected and observed 
events, and the measured cross section as a function of $b^\prime$ mass
in the $2\gamma + 2$ jets channel.  The integrated luminosity
is 79 pb$^{-1}$.}
\begin{tabular}{c|c|ccc|cc}
 && \multicolumn{3}{c|}{Events} & 
   \multicolumn{2}{c}{$\sigma_{b^\prime\bar{b^\prime}}\times 
  B(b^\prime\bar b^\prime \to \gamma\gamma b\bar b)$ (pb)}\\
\cline{3-7}
$m_{b^\prime}$  & Acceptance
&& Expected & Expected && Upper limit \\
(GeV/c$^2$) & (\%) & Observed & Signal ($B = 1.6$\%)   & 
Background & Value & (95\% CL) \\
\hline
\hline
50 & $2.76\pm0.40$ & 20 & $126\pm20$ & $15.5\pm2.3$ & $2.03\pm2.33$ 
& $6.11$ \\
60 & $5.31\pm0.71$ & 18 & $101\pm15$  & $14.1\pm2.1$ & $0.91\pm1.14$ 
& $2.91$ \\
70 & $8.19\pm1.08$ & 15 & $68.3\pm9.7$ & $11.0\pm1.7$ & $0.61\pm0.66$ 
& $1.76$ \\
80 & $10.45\pm1.37$ & 11 & $42.9\pm6.1$ & $8.4\pm1.3$ & $0.31\pm0.44$ 
& $1.08$ \\
90 & $11.90\pm1.52$ & 8 & $26.3\pm3.6$ & $6.2\pm1.0$ & $0.18\pm0.32$ 
& $0.76$ \\
100 & $13.23\pm1.68$ & 6 & $16.5\pm2.3$ & $4.4\pm0.8$ & $0.15\pm0.25$ 
& $0.59$ \\
120 & $15.73\pm2.00$ & 3 & $7.5\pm1.0$ & $2.4\pm0.5$ & $0.05\pm0.15$ 
& $0.32$ \\
140 & $16.28\pm2.06$ & 3 & $3.4\pm0.5$ & $1.5\pm0.4$ & $0.11\pm0.14$ 
& $0.36$ \\
\end{tabular}
\end{table*}

\end{document}